\documentclass{article}
\usepackage{graphicx} % Required for inserting images
\usepackage{tabularx}
\usepackage{url}
\usepackage[margin=1in]{geometry}
\usepackage{float}
\usepackage{xurl}
\usepackage{longtable}
\usepackage{array}
\usepackage{xcolor}
\usepackage{caption}
\usepackage[numbers,sort&compress]{natbib}
\usepackage{hyperref}

\usepackage{authblk}

\title{Comparing the Performance of Leading VQE Algorithms for Computing Ground-State Energies of Amino Acids}

\author[1]{Sanskriti Shindadkar}
\author[2,3]{Clyde Villacrusis}
\author[4]{Jasper Andrews}
\author[2]{Brandon Yan}

\affil[1]{Department of Bioengineering, UCLA Samueli School of Engineering}
\affil[2]{Department of Computer Science, UCLA Samueli School of Engineering}
\affil[3]{Department of Linguistics, UCLA College of Letters and Science}
\affil[4]{Department of Mathematics, UCLA College of Letters and Science}

\date{July 1, 2026}

\begin{document}

\maketitle
%%%%%%%%%%%%%%%%%%%%%%%%%%%%%%%%%%%%%%%%%%%%%%%%%%%%%%%%%
\begin{abstract}
Simulating molecules is a major application of quantum computing, with the potential to overcome exponential scaling constraints of classical computation. Researchers use different methods in order to evaluate the readiness of NISQ computers in order to test current simulation capabilities. We present an integrated repository with reproducible benchmarks of over 10 different ansatz from published papers and two different truncation methods, applicable to any set of mapped hamiltonians, providing a single pipeline for comparing performance along multiple axes, including variance and computational time, among others. We apply them to simulate different amino acids, using hamiltonians taken from the QMProt Dataset. We then ran four separate experiments. First, we quantified noise resilience by optimizing the same hardware-efficient ansatz under identical initialization while sweeping PennyLane noise channels and strengths, and measuring parameter drift, cosine similarity of optimal parameters, and energies evaluated on noiseless versus noisy backends. We then studied barren-plateau–related trainability via gradient-variance diagnostics and optimization trajectories across initialization strategies and ansatz depth on small systems. We then compared adaptive versus fixed ansätze at matched parameter budgets, reporting outer-loop iterations, wall time, and especially total cost-function evaluations to fairly contrast greedy adaptive growth with layered hardware-efficient circuits. Lastly, we mapped accuracy versus expressive capacity by sweeping the number of retained adaptive operators and recording ground-state energy error relative to classical references.

\end{abstract}
%%%%%%%%%%%%%%%%%%%%%%%%%%%%%%%%%%%%%%%%%%%%%%%%%%%%%%%%%

%%%%%%%%%%%%%%%%%%%%%%%%%%%%%%%%%%%%%%%%%%%%%%%%%%%%%%%%%
\section{Introduction}

The accurate simulation of molecular systems remains one of the most formidable challenges in computational science. At the heart of this challenge lies the Electronic Structure Problem (ESP): solving the Schrödinger equation to determine the ground-state energy (GSE) of a many-electron system. Classical methods, while sophisticated, face an exponential scaling wall; as the molecular system size increases, the number of possible electronic configurations grows combinatorially, rendering exact solutions for large biological molecules computationally intractable. While chemical accuracy (about 4.184 kJ/mol) is enough for many molecular calculations, some scientific questions require higher accuracy. For instance, reliable polymorph ranking can require sub-chemical accuracy (a few tenths of a kJ/mol)\cite{ludik2024polymorphism}, making greater accuracy of great interest for pharmaceutical developers. 
\newline
\newline
A variety of classical Density Functional Theory (DFT) is being used for for many molecular ground-state energy calculations and datasets. Ramakrishnan et al. ~\cite{ramakrishnan2014quantum} computed equilibrium geometries and many properties, including free energy for about 134k stable CHONF molecules using the B3LYP density functional with the 6-31G basis set, making a reference dataset ~\cite{ramakrishnan2014quantum}. Later, Narayanan et al. ~\cite{narayanan2019accurate} computed G4MP2 energies for about 133,000 QM9 organic molecules and benchmarked G4MP2 enthalpies of formation against 459 molecules with an accuracy of 0.79 kcal/mol for G4MP2 enthalpies of formation ~\cite{narayanan2019accurate}. 
% \newline
% \newline
Researchers are also beginning to explore Large Language Models  (LLMs) for molecular and materials property prediction. For crystalline materials, Niyongabo Rubungo et al. \cite{niyongabo2025llmprop} proposed LLM-Prop, a T5-based model that predicts crystal properties from text descriptions and reported competitive performance on several crystal-property tasks \cite{niyongabo2025llmprop}. However, current results for direct molecular property prediction using general-purpose LLMs is very mixed. Busch et al. \cite{busch2026incontext} evaluated GPT-4.1, GPT-5, and Gemini 2.5 variants on MoleculeNet tasks including QM7 atomization energy, distinguishing in-context learning from memorization using progressive molecular blinding to distinguish in-context learning from memorization \cite{busch2026incontext}. In their QM7 analysis, they reported that predictions were ''highly inaccurate'' and ''common errors include[d] predictions at the wrong scale (around 100) and with the wrong sign.''
\newline
\newline
Overall, classical methods, such as DFT, still generally dominate over LLM-based methods. However, exact classical electronic-structure methods are constrained by the cost of representing many-electron wavefunctions explicitly. Quantum methods, on the other hand, can represent wavefunctions in a Hilbert space that scales exponentially with the number of qubits. A qubit lives in a two-dimensional Hilbert space, and that n qubits are described by a vector in a $2^n$-dimensional Hilbert space ~\cite{mcardle2020quantum}. This matches the structure of many-electron quantum mechanics since qubits can represent quantum states using superposition and entanglement rather than explicitly storing all electronic configurations classically~\cite{cao2019quantum,kassal2011simulating}. Furthermore, where classical approaches struggle with strongly correlated systems as they necessitate multi-configurational electronic wavefunctions~\cite{vaquero2024physically}, qubits are capable of representing these systems and then approximating the ground state of the Hamiltonian variationally using a parameterized trial state~\cite{huang2022simulating}. 
\newline
\newline
In the current Noisy Intermediate-Scale Quantum (NISQ) era, traditional algorithms like Quantum Phase Estimation (QPE) remain impractical due to their requirement for high-depth circuits and fault-tolerant error correction ~\cite{preskill2018nisq}. To circumvent these hardware limitations, the Variational Quantum Eigensolver (VQE) was introduced as a hybrid quantum-classical algorithm~\cite{peruzzo2014variational}. VQE leverages the variational principle, which states that the expectation value of the Hamiltonian H for any trial wavefunction is an upper bound to the true ground-state energy E0:
\[
E(\theta) =
\frac{\langle \psi(\theta) | H | \psi(\theta) \rangle}
{\langle \psi(\theta) | \psi(\theta) \rangle}
\geq E_0
\]
\newline
\newline
By offloading the parameter optimization to a classical optimizer and using the quantum processor solely for state preparation and measurement, VQE significantly reduces the coherence time requirements~\cite{mcclean2016theory}. Since its inception, VQE has been successfully applied to small molecules such as \(\mathrm{H_2}\), LiH, and \(\mathrm{BeH_2}\) ~\cite{kandala2017hardware}.
In drug discovery, lead-compound optimization often depends on accurate estimates of protein--ligand binding free energies. Quantum-mechanical methods are increasingly used to improve the treatment of electronic interactions in these calculations, motivating more scalable electronic-structure workflows~\cite{cavasotto2020binding}.
% \newline
% \newline
Understanding the energy landscapes of amino acids and their transition states is vital for modeling enzymatic catalysis and protein folding. Precise calculations of molecular interactions are important for capturing non-covalent effects, electrostatics, and protein-protein interactions, while classical force fields can remain limited by fixed functional forms and transferability constraints~\cite{lopes2015current,piana2020forcefield}. However, the biological scale introduces high-dimensional Hamiltonians, necessitating VQE algorithms that are not only accurate but also computationally efficient in terms of measurement overhead and circuit depth.
\newline
\newline
Furthermore, recent developments in VQEs suggest promising results in protein folding simulation, potentially opening opportunities for drug discovery against disordered proteins long considered difficult to target~\cite{uttarkar2026qupepfold}. 
The rapid proliferation of VQE variants has created a ''zoo'' of algorithms, including to what is known as a chemically-inspired ansatz like the Unitary Coupled Cluster (UCCSD), which provide high accuracy but often result in prohibitively deep circuits ~\cite{anand2022quantum}.
Hardware-Efficient Ansatz (HEA), which prioritize low circuit depth at the cost of being susceptible to ''Barren Plateaus'' during optimization~\cite{mcclean2018barren}.
Adaptive Ansatz, such as ADAPT-VQE, which grows the operator pool dynamically to balance accuracy and efficiency~\cite{grimsley2019adaptive}.
Despite this innovation, the field lacks a unified metric for comparison. Most studies benchmark algorithms on different hardware, using different optimizers, or on varying molecular geometries. This fragmentation makes it difficult for researchers to select the optimal algorithm for applications involving activation or transition-state energetics, where accurate energy differences are essential to reliable mechanistic modeling~\cite{abidi2022transitionstates}.
To address these challenges, this study utilizes the newly released QMProt database~\cite{coronas2025qmprot}. Unlike previous datasets focused on small organic molecules (e.g., QM9), QMProt provides the mapped Hamiltonians specifically for the core amino acids, providing a standardized ''battleground'' for quantum algorithms in a biological context. We present a comprehensive framework designed to evaluate VQE performance across three critical axes:
\begin{itemize}
    \item Accuracy: Proximity to the exact diagonalization, or FCI, limit within chemical accuracy, defined as 1.6 mHa.
    \item Computational speed: Number of iterations and total shots required to reach convergence.
    \item Error spread: Robustness of the algorithm against statistical noise and shot-noise fluctuations.
\end{itemize}
By applying this framework to the amino acids within QMProt, we aim to provide a roadmap for the scalable simulation of larger biomolecules, ultimately bridging the gap between quantum theory and practical pharmaceutical application.
%%%%%%%%%%%%%%%%%%%%%%%%%%%%%%%%%%%%%%%%%%%%%%%%%%%%%%%%%

%%%%%%%%%%%%%%%%%%%%%%%%%%%%%%%%%%%%%%%%%%%%%%%%%%%%%%%%%
\section{Methodology}

\begin{figure}[H]
\centering
\includegraphics[width=0.95\textwidth]{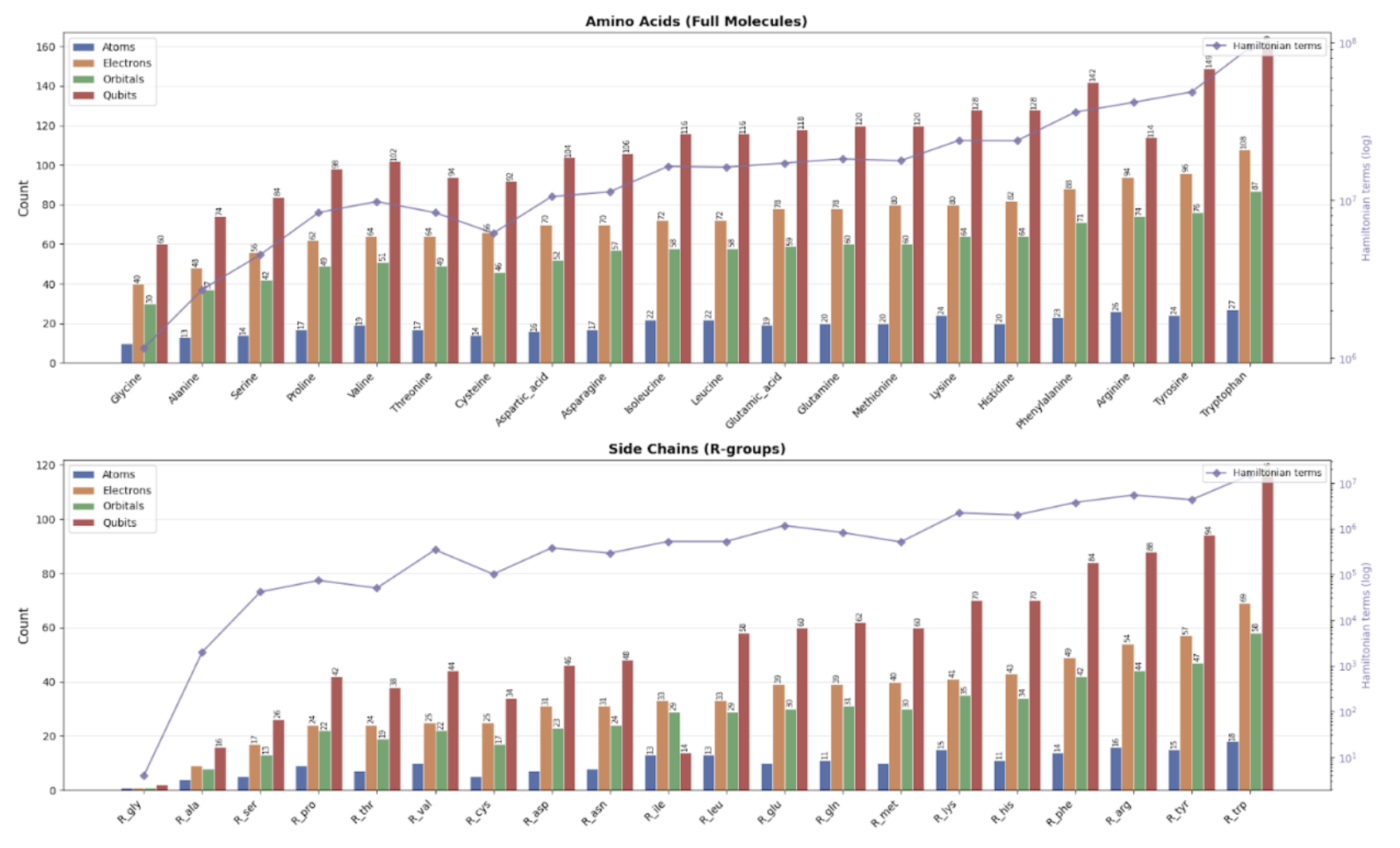}
\caption{Number of atoms, electrons, orbitals, and qubits for the molecules available in the QMProt dataset.}
\label{fig:qmprot-dataset-summary}
\end{figure}

We exclusively used the QMProt Dataset. They compute Hamiltonians using OpenFermion with inputs including coordinates, charge, spin (via multiplicity), and a basis set, then run PySCF with SCF, and finally output a fermionic Hamiltonian using Jordan-Wigner. Note that PySCF is a modular open-source computational chemistry suite designed for electronic structure calculations, including Hartree-Fock. There are numerous other options for computing fermionic Hamiltonians, such as Bravyi-Kitaev or generalized encodings. Jordan-Wigner Hamiltonians map each orbital to a qubit and this mapping structure is deterministic and interpretable. Bravyi-Kitaev instead stores information about orbital occupation across multiple qubits using a binary tree structure, reducing circuit depth and noise accumulation, especially for larger subsystems. However, the advantage is minimal for small systems like amino acids. The choice of fermion-to-qubit mapping can significantly affect circuit resources, and is a potential research direction for more explicit molecular benchmarking.
\cite{tilly2022vqe}. They then extracted the 3D atomic coordinates from PubChem SDF files. 
\newline
\newline
There are 45 molecules in the dataset, including water, the amino group, the main amino acids, and their radicals. A distribution of their reference ground state energies, as well as qubit mappings, and electron representations can be seen in Figure~\ref{fig:qmprot-dataset-summary}. The energies of the amino acids in the dataset ranged from around -300 to -780 Hartree.

There have been dozens of different VQE algorithms developed by researchers over the last few years. For this repository, we focused on major ones present in the academic and industry communities. A list of them can be found in  Table~\ref{tab:vqe-methods}. They include VQEs from Pennylane, recently published papers, and versions of ADAPT-VQE and Hardware-Efficient VQEs. We used available code from the researchers’ GitHubs, and adapted them to run on the Pennylane/Qiskit ecosystem so they could all be run from our centralized repository. 
\newline
\newline
This work benchmarks VQE variants under an explicit qubit budget with two possible truncation methods implemented. The goal is to preserve the highest energetic contributions while reducing circuit width and measurement cost, enabling consistent comparisons across algorithms on NISQ-relevant problem sizes. Many quantum chemistry Hamiltonians, once mapped to qubits, contain a large number of Pauli strings with small coefficients. For example, phenylalanine from the dataset is default mapped to 132 qubits. Using hard cutoffs on coefficient magnitude can help reduce to a substantially smaller effective Hamiltonian while preserving the main optimization landscape~\cite{preskill2018nisq}.
\newline
\newline
Near-term benchmarking is constrained by limited qubit counts and the depth of circuits that current hardware can reliably execute, to such an extent that directly representing large molecular systems is infeasible for the time being~\cite{trivino2023complete}. However, by reducing the number of qubits required to map a system, we’re reducing our problems to a complexity that we can represent on current hardware. A common strategy is to reduce the effective qubit space by focusing on the most chemically relevant orbitals. Another approach is to take advantage of symmetries in fermionic Hamiltonians to eliminate redundancies and lower qubit requirements without altering the problem~\cite{moll2018quantum}. Regardless, fidelity is also important, and through methods such as extrapolation to the zero noise limit, the effects of errors and decoherence can be mitigated~\cite{temme2017error}. Such reductions, together with validity-preserving schemes, enable VQEs to converge on near-chemical accuracy using circuits shallow enough for NISQ devices.

\subsection{Active Space Truncation}
To enable benchmarking under tight qubit constraints to allow simulation, we used a systematic active-space truncation protocol to reduce the electronic Hamiltonian prior to qubit mapping. All electronic structure calculations were performed using PySCF. For each molecule, we first computed Hartree-Fock molecular orbitals and orbital energies. Core orbitals were identified and frozen based on energetic separation and chemical considerations (standard from the library). Candidate active orbitals were proposed using a combination of orbital energy proximity to the Fermi level, natural orbital occupation diagnostics from correlated calculations, and automated selection strategies such as the Atomic Valence Active Space (AVAS) method implemented in PySCF~\cite{guo2021vqe_ucc}. This procedure defined an active space characterized by a specified number of active electrons and spatial orbitals.
\newline
\newline
The active space was validated using multi-configurational calculations within PySCF. Complete active-space configuration interaction (CASCI) or complete active-space self-consistent field (CASSCF) calculations were performed to assess whether the selected orbitals captured the dominant static correlation effects~\cite{sun2017general,pyscf_mcscf_docs}. Validation criteria included the presence of significant configuration mixing and fractional natural occupation numbers within the active subspace. 
\newline
\newline
After validation, a reduced second-quantized Hamiltonian was constructed in the active orbital basis. Frozen-core and external orbitals were excluded, yielding a fermionic Hamiltonian restricted to the active space. This reduced Hamiltonian was generated using either Qiskit Nature’s active-space transformation utilities or the OpenFermion–PySCF interface. The resulting fermionic operator was then mapped to qubits via standard fermion-to-qubit transformations. The final qubit count is twice the number of spatial active orbitals, making sure that qubit reduction occurred at the physically meaningful fermionic level before any of the VQEs.
\newline
\newline
The mathematical background for this is as follows:
\begin{itemize}

    \item The pipeline starts with the full Hamiltonian input.

    \item HF calculated for an orbital basis and a reference occupancy pattern.

    \item MP2 diagnoses which orbitals are no longer ``cleanly'' 0 or 2 occupied, using a perturbative method around a single-reference HF state.

    \item The one-particle reduced density matrix, 1-RDM, is calculated and diagonalized.

    \item Correlation score highest for around 1, or half occupied.

    \item Orbitals are partitioned. Frozen core, for doubly-occupied orbitals, assumes certain low-energy orbitals remain doubly occupied in every determinant considered.

    \item Active space electrons calculated.

    \item Validating that reduced model with CASCI/CASSCF.

    \item Mapping the resulting fermionic Hamiltonian to qubits with Jordan-Wigner, for consistency with the original paper.
\end{itemize}

\begin{figure}[H]
\centering
\includegraphics[width=0.85\textwidth]{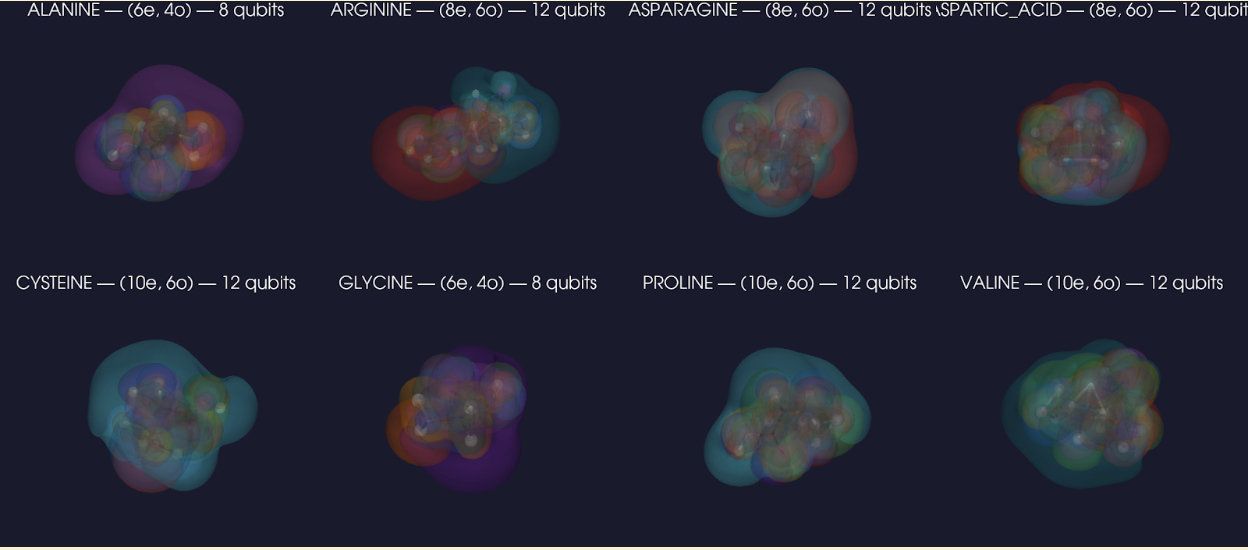}
\caption{Active orbitals of model amino acids, visualized.}
\label{fig:active-space-orbitals}
\end{figure}

In our pipeline, we’ve also implemented functions that allow researchers to visualize the chosen active spaces in an interactive manner. Figure~\ref{fig:active-space-orbitals} displays the active spaces chosen for a few model amino acids.

We also have a flag to enable constructing a reduced (truncated) qubit Hamiltonian that retains only a subset of Pauli terms from the full mapped Hamiltonian.

\subsection{Contextual Subspace Reduction}

To mitigate the high qubit requirements associated with amino acid Hamiltonians, we implemented an optional Contextual Subspace (CS) reduction step. Based on the theory proposed by Kirby et al., this method partitions the Hamiltonian (H) into a classical component and a quantum component. This was integrated into the repository using code inspired from the repository used in the paper~\cite{kirby2021contextual}.
\newline
\newline
The reduction process follows a three-step workflow:
\begin{enumerate}
    \item Sub-Hamiltonian extraction: We identify the largest noncontextual sub-Hamiltonian, $H_{\mathrm{NC}}$, which consists of a subset of Pauli terms whose expectation values can be simultaneously determined by a classical hidden-variable model. This extraction uses a greedy depth-first search, DFS, prioritized by the magnitude of the coefficients.

    \item Classical optimization: The ground state of $H_{\mathrm{NC}}$ is solved classically. We use a hybrid optimization strategy with exhaustive enumeration of all $2^m$ sign assignments for the commuting sector and differential evolution for the anti-commuting sector.

    \item Qubit projection: The classical solution provides $m$ stabilizer constraints. We apply a sequence of Clifford rotations to map these constraints onto single-qubit $Z$ operators. These stabilized qubits are then projected out, resulting in a reduced effective Hamiltonian acting on as few as $n - m$ qubits, for example, reducing a 10-qubit water Hamiltonian to 4 qubits. Intermediate qubit targets are supported by releasing generators back to the quantum simulation.
\end{enumerate}
\subsection{Basis Transformation and State Preparation}

The Clifford rotations used in CS reduction shift the system into a noncontextual qubit basis. This necessitates a corresponding transformation of the Hartree-Fock (HF) initial state to ensure physical consistency.
\newline
\newline
Our framework utilizes a \texttt{\_prepare\_initial\_state()} function that applies the identical unitary transformation used in the Hamiltonian reduction to the initial $|HF\rangle$ state. This transformed state is then projected onto the reduced qubit subspace. This ensures that the VQE optimization begins at a chemically relevant starting point, even within the transformed Hilbert space.

\subsection{Framework Integration}

A core feature of our framework is its ability to handle different VQE variants in the reduced subspace without manual reconfiguration. Algorithms are categorized into two primary logic flows:

\begin{itemize}
    \item Hardware-Native Algorithms: Methods such as Hardware-Efficient VQE and QAOA-inspired VQE are treated as fully native. These depend only on the number of qubits and do not rely on an assumed fermionic structure, allowing them to operate directly on the reduced Hamiltonian.

    \item Fermionic-Adapted Algorithms: For algorithms that typically assume an orbital structure (e.g., Vanilla VQE, ADAPT-VQE, IQCC), the framework clears \texttt{n\_electrons} and \texttt{n\_orbitals} as a safeguard, since these quantities are undefined in the rotated basis.
\end{itemize}

In the case of Qubit-ADAPT, the standard ''occupied-to-virtual'' operator pool becomes physically meaningless after the Clifford basis rotation. To maintain expressivity, our framework automatically switches the operator pool to a full Pauli pool whenever CS reduction is active. This allows the adaptive algorithm to explore the transformed space effectively despite the loss of traditional molecular orbital structure.
\newline
\newline
One strategy we used to ensure that our VQE solutions for GSE were accurate is by making sure the reference state outputted the correct Hartree Fock energy, and setting all parameters to zero at the HF. We verified correctness of our variational implementations by computing the Hartree–Fock reference energy ⟨HF|H|HF⟩ and confirming that optimized VQE energies were not greater than this value (in noiseless simulations). Because the Hartree–Fock state is contained within the ansatz parameter space, the variational principle guarantees that the optimized energy is bounded above by the Hartree–Fock energy. 

\subsection{Experiments Run}
We ran four different categories of experiments; adding in additional algorithms to the repository using the current framework will allow them to be run again. These experiments are insightful for future runs on hardware. For instance, although there already exists research on ‘noise resilience’ of different VQE algorithms, there aren’t clear benchmarks that compare performance for biologically significant molecules.  
\begin{itemize}
    \item Noise resilience: Same molecule, same ansatz depth and shared initial parameters; optimized with \texttt{hardware\_efficient\_vqe} and PennyLane noise models while sweeping noise channel and strength $p$. Comparing noisy optimum vs noiseless baseline on energy and parameter drift ($L_2$, cosine); optional batch runs across molecules with aggregated heatmaps.

    \item Barren plateaus / initialization: Hardware-efficient ansatz at several depths:
    \begin{itemize}
        \item Part A: gradient variance or mean vs gradient vs depth over random initializations and named initial strategies
        \item Part B: repeated Bayesian optimization runs comparing convergence and final energy across strategies at fixed depth. Initially, we framed the runs as with COBYLA at first as a standard VQE baseline, but we additionally ran the 24 molecules with the Bayesian optimization with 5 trials per initialization at fixed depth. As a result, near-identity and small-random initializations generally outperforms random-uniform in the final energy and convergence behavior, as can be seen in Figure ~\ref{fig:part_b_convergence_summary_readable}.
    \end{itemize}

    \item Trainability (matched parameters): For each target parameter count, e.g. 10 / 50 / 100, ran \texttt{qubit\_adapt\_vqe} fixed max operators vs \texttt{hardware\_efficient\_vqe} with depth chosen to match $\geq$ that many parameters. Recorded algorithm-reported iterations, wall time, final energy vs CASCI, and total cost-function evaluations wrapped counter for a fair trainability comparison.

    \item Accuracy vs number of parameters: \texttt{qubit\_adapt\_vqe} with a modular $k$-list: one run up to $\max(k)$ greedy operators; sliced energies at each $k$ vs HF/CASCI, plus cumulative cost-eval and runtime snapshots. Note that by default, the algorithms are using active-space Hamiltonians without CS so ADAPT can grow from HF CS optional where noted in docs.
\end{itemize}

\section{Results}
\subsection{Reference Energy Verification}
First, we verified that the reference energies matched the hartree-fock energies, and then the VQEs were able to at least equal to the hartree-fock energies or a lower value. This was successfully done. Several of the results for different molecules can be seen in Figure 3.

\begin{figure}[H]
\centering
\includegraphics[width=0.95\textwidth]{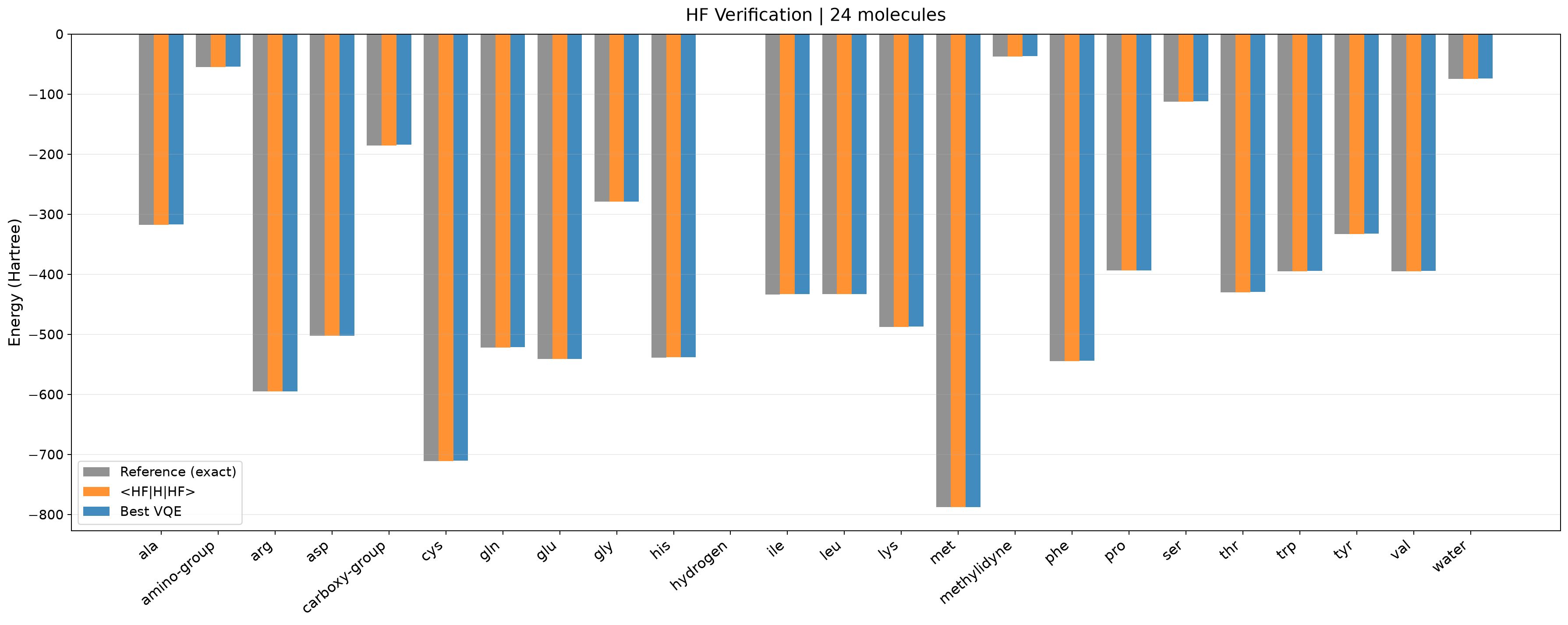}
\caption{HF Verification VS other metrics on the 24 QMProt molecules. Bayesian Optimizations was also used in this graph.}
\label{fig:hf-verification}
\end{figure}

\subsection{Experiment 1: Influence of Initialization Strategies on Convergence Speed and Energies}
\begin{figure}[H]
\centering
\includegraphics[width=0.95\textwidth]{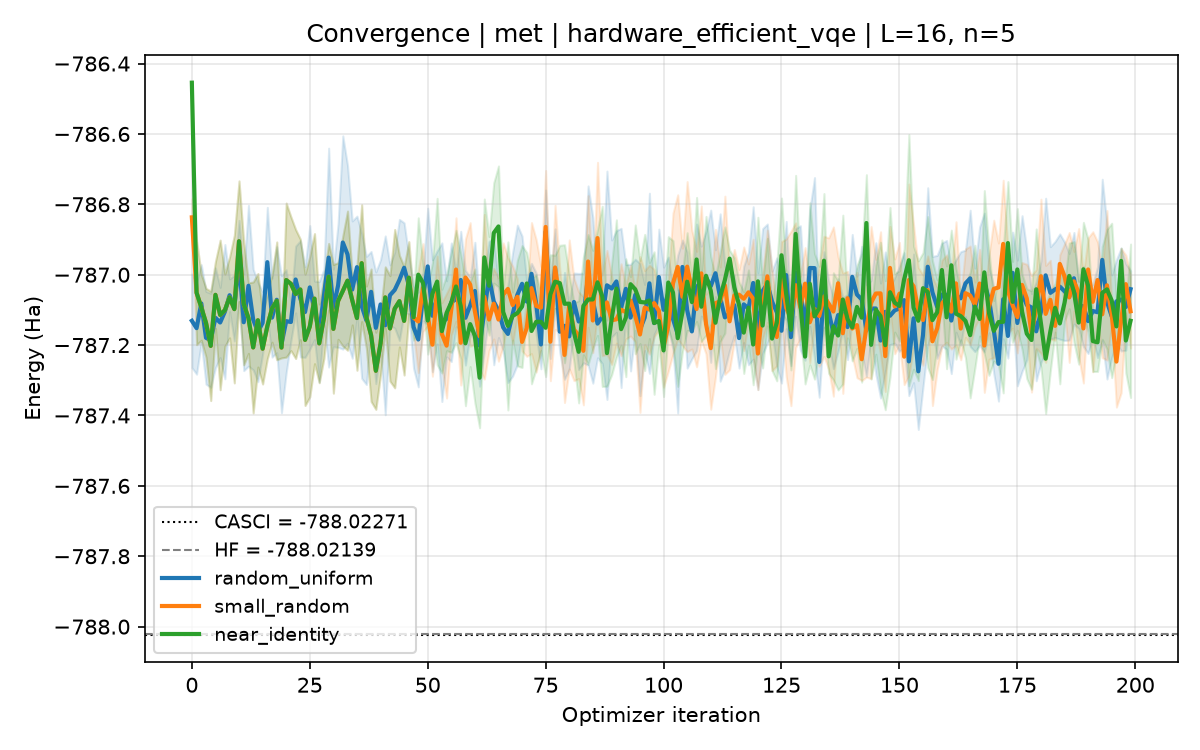}
\caption{In the Methionine example, all three initialization strategies quickly settle into similar oscillatory plateaus around -787.1 Ha and stay above the CASCI/HF reference (-788.02 Ha), showing that the near-identity and small-random dominates the random-uniform energy.}
\label{fig:onvergence_by_init_met}
\end{figure}

\begin{figure}[H]
\centering
\includegraphics[width=0.95\textwidth]{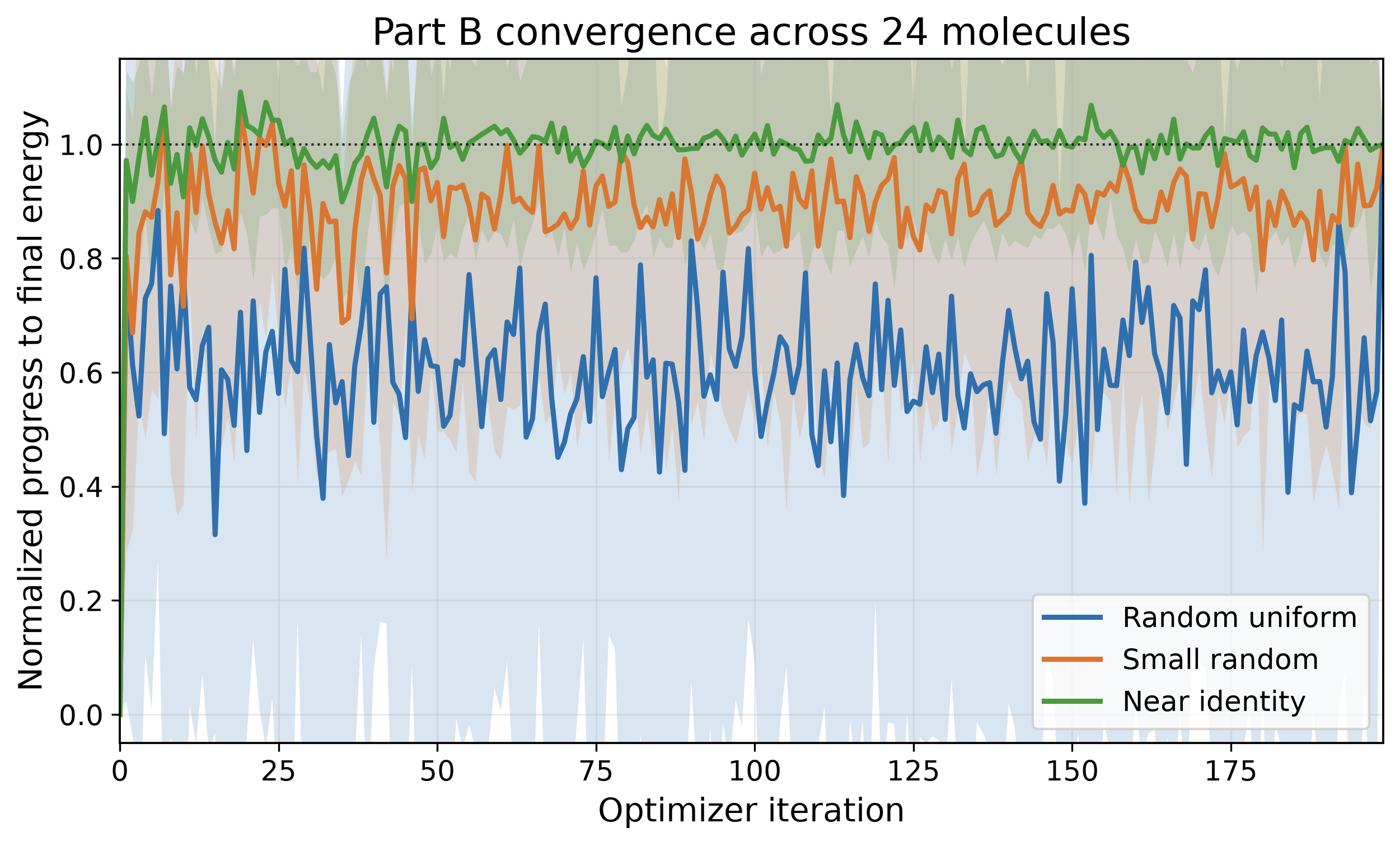}
\caption{This panel figure presents per-molecule convergence-by-initialization curves, enabling side-by-side inspection of trainability patterns across the 24 molecules. The general trend is that the energies converge around the 2-5 optimizer iteration, and then fluctuate after that, regardless of initialization method. Note that these runs do not have any noise strengths in them.}
\label{fig:part_b_convergence_summary_readable}
\end{figure}

\begin{figure}[H]
\centering
\includegraphics[width=0.85\textwidth]{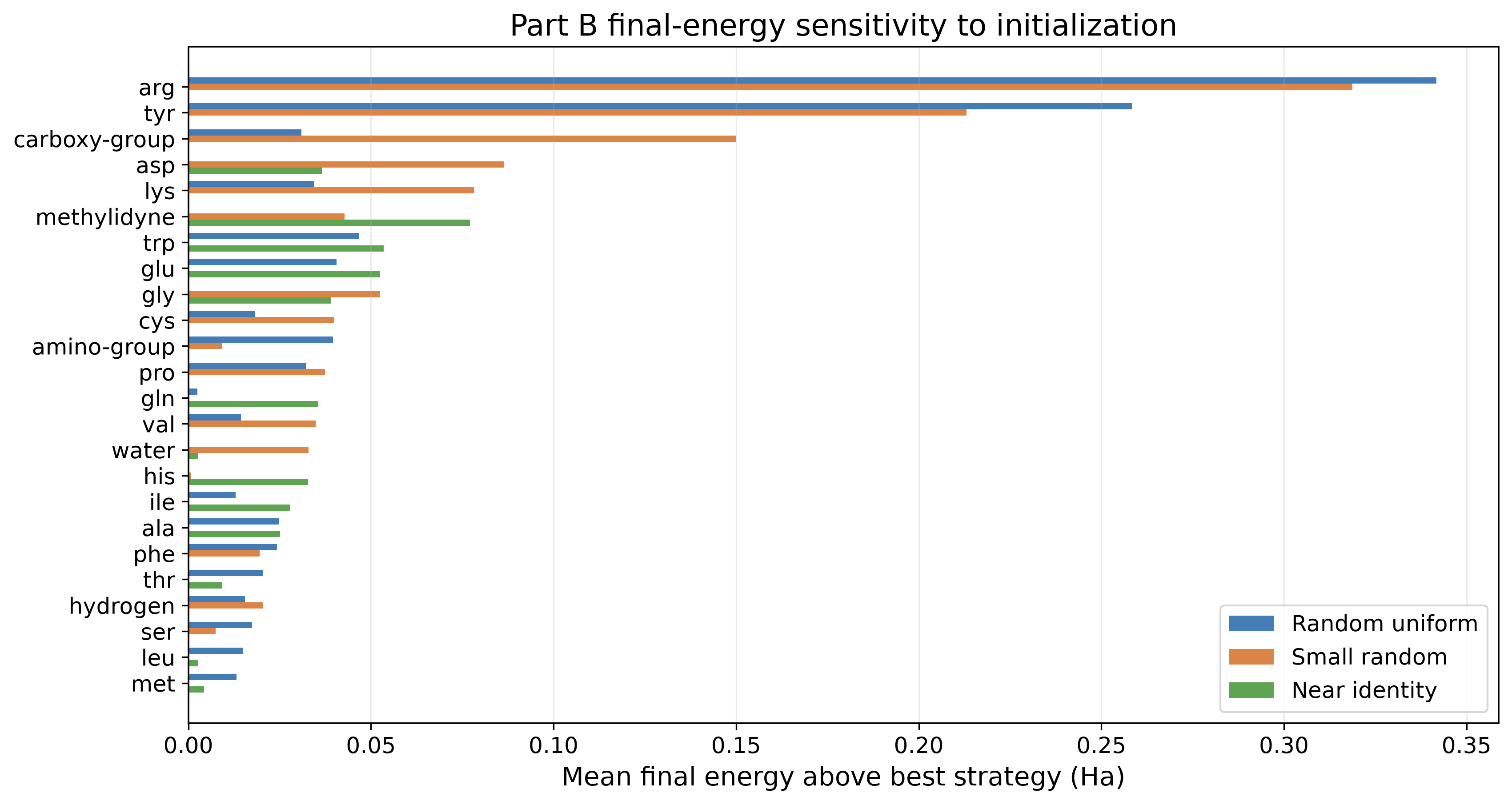}
\caption{This panel figure summarizes per-molecule final-energy-by-initialization comparisons, showing which initialization strategy yields the best endpoint for each molecule. There is no statistically significant trend in variation of final energies across different initialization strategies. Note that these runs do not have any noise strengths in them.}
\label{fig:part_b_final_energy_delta_readable}
\end{figure}

As can be seen in Figure~\ref{fig:onvergence_by_init_met}, convergence for methionine using hardware efficient VQE happens very quickly (around the 2nd/3rd iteration), regardless of the initialization strategy. Oscillation is generally on the order of 0.4 Hartree, which is about 1.7439e-18 J. This trend continues for the other molecules in the dataset, as shown in Figure~\ref{fig:part_b_convergence_summary_readable}. Note that this run do not have any noise strengths in them. The general trend of minimal differences in the initialization strategies for convergence speed continues in regards to influence on final energy, as is shown in Figure~\ref{fig:part_b_final_energy_delta_readable}

\begin{figure}[H]
\centering
\includegraphics[width=0.95\textwidth]{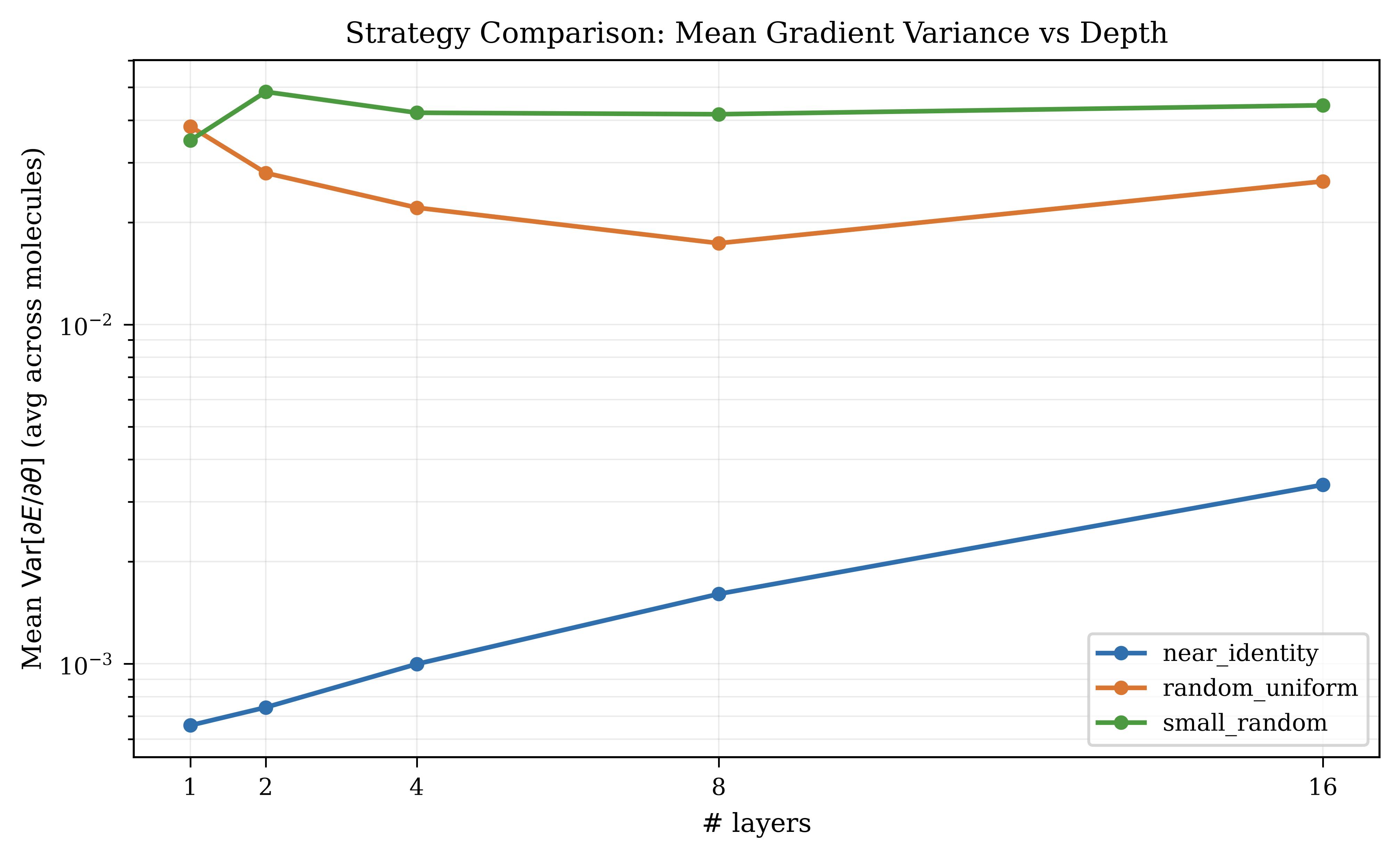}
\caption{Comparing the mean variance across molecules, compared to the number of layers, for the three different strategies. Using near-identity consistently results in the least variance, though the variance does increase with the number of layers, whereas random-uniform and small-random average greater variance, though less penalized by greater layer counts.}
\label{fig:mean_gradient_variance_vs_depth_all_molecules}
\end{figure}

\subsection{Experiment 2: Effects of Noise on Parameter Drift}

\begin{figure}[H]
\centering
\includegraphics[width=0.95\textwidth]{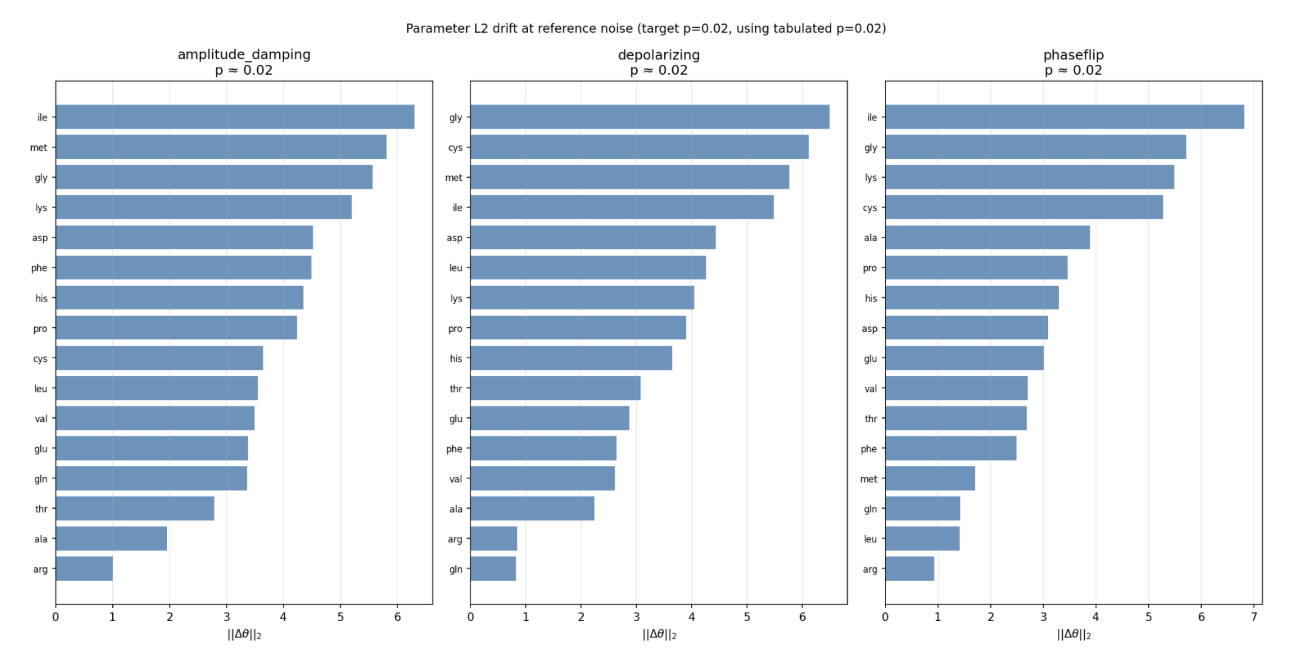}
\caption{Parameter L2 Drift at Reference Noise. These, on their own, don’t tell you much about how significant the parameter drift is relative to baseline, but does tell you about relative drift between molecules. Refer to Figure~\ref{fig:l2-heatmap} for further information. These were run on 36 parameters on 6-qubit, 2-layer RY hardware-efficient setup; an L2 norm of roughly 6 is compatible with “order-1 radian” average coordinate shifts. This was generated from experiment 1.}
\label{fig:param-l2-drift}
\end{figure}

\begin{figure}[H]
\centering
\includegraphics[width=0.95\textwidth]{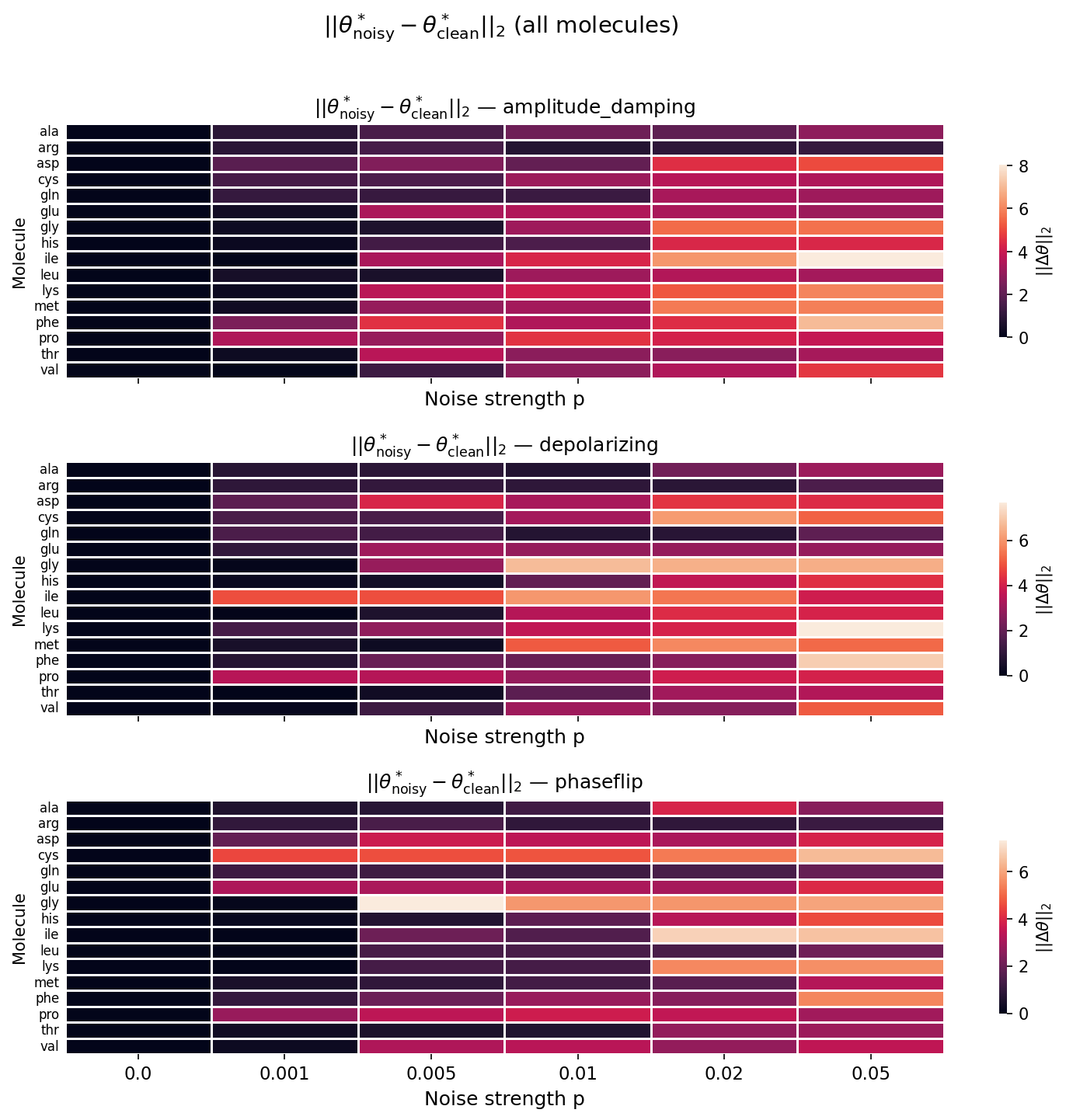}
\caption{Heatmap of parameter drift measured using the L2 norm. Notice how when the noise strength p increases, the parameter drift has a general trend of increasing as well.}
\label{fig:l2-heatmap}
\end{figure}

Figures ~\ref{fig:param-l2-drift} and ~\ref{fig:l2-heatmap} use L2 distance as a metric to talk about parameter drift at reference noise. L2 distance, also known as Euclidean distance, refers to the straight-line distance between two points; in our context, it refers to the distance between initial and final energy and parameter values. Cosine similarity measures the cosine of the angle between two vectors. In our context, cosine similarity helps compare how energy and parameter drift change between our noisy optimum and noiseless baseline. The cosine similarity heatmap can be found as Figure {insert} in the appendix, and is in line with what is presented in the L2 heatmap, which is Figure ~\ref{fig:l2-heatmap} .

\subsection{Experiment 3: Investigating Energy Error as a Function of Parameters}
\begin{figure}[H]
\centering
\includegraphics[width=0.95\textwidth]{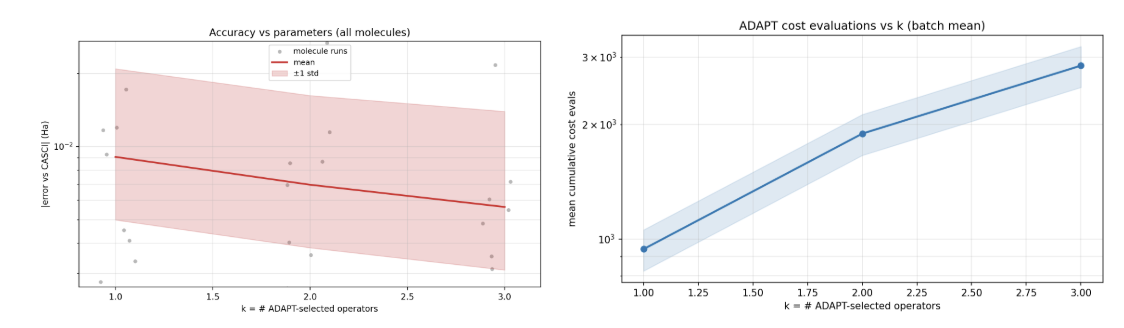}
\caption{Investigating energy error vs n\_parameters for Adapt-VQE, compared to the CASCI. All of the amino acids in the QMProt dataset were run up to $k = 3$. Larger experiments can be run. Error decreases as n-parameters increases, though cost increases.}
\label{fig:adapt-energy-error}
\end{figure}

\begin{figure}[H]
\centering
\includegraphics[width=0.95\textwidth]{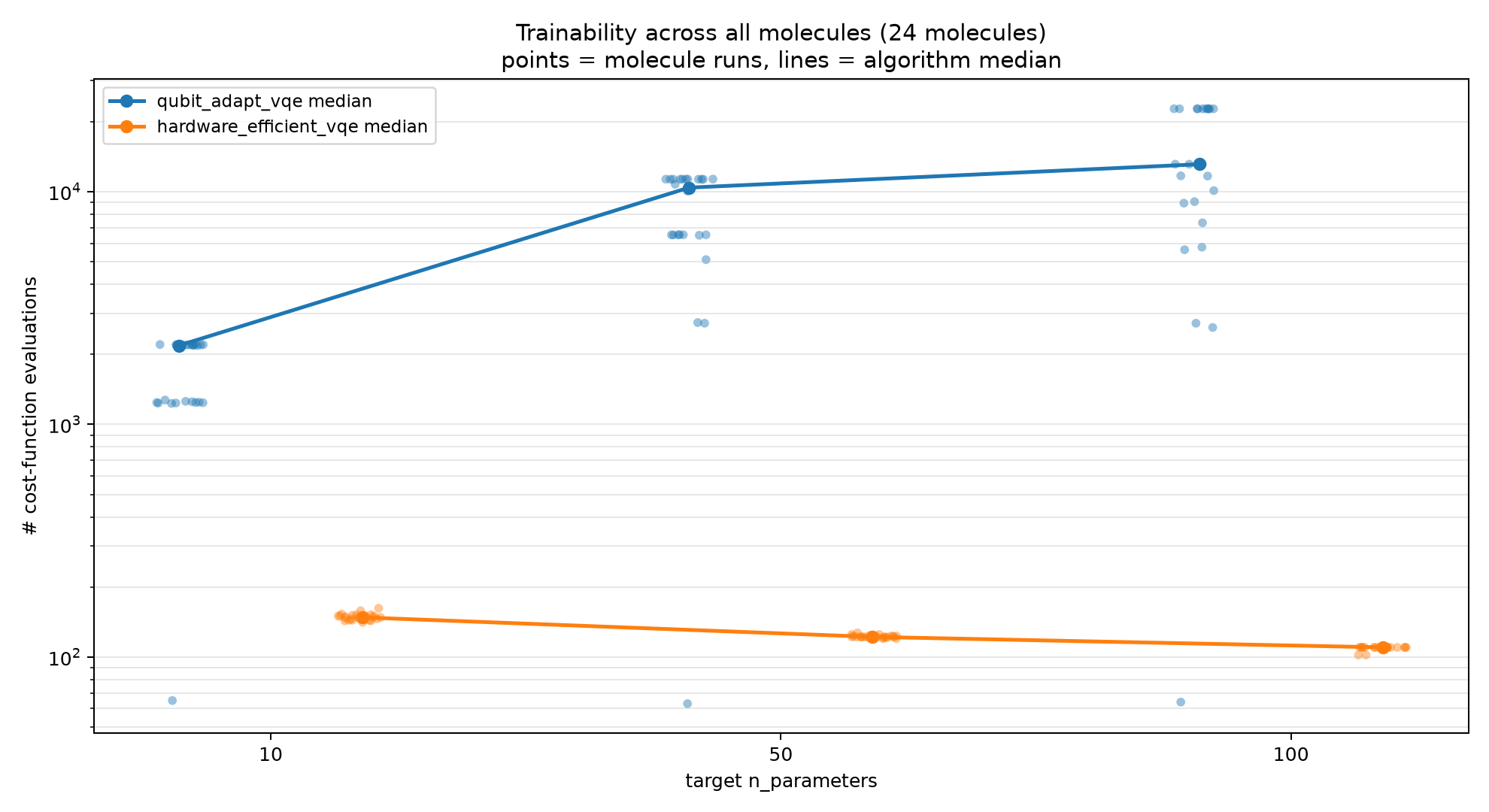}
\caption{Investigating cost-function evals for all molecules between qubit-ADAPT and hardware-efficient VQE. The hardware-efficient VQE is consistently cheaper by a magnitude of 10-20 times than qubit-ADAPT.}
\label{fig:trainability-fnevals}
\end{figure}

Running just from k = 1 to k=3, Figure ~\ref{fig:adapt-energy-error} shows how increasing the ADAPT-selected operators can decrease the energy error, while cost evaluations go down.
\newline
\newline
Figure ~\ref{fig:trainability-fnevals} shows the general trend across molecules for which increaseing the number of parameters from under 10, to 50, to 100 increases the number of cost-function evaluations on the qubit-adapt vqe, but actually lowers it on the hardware adapt VQE. A molecule by molecule breakdown can be found in the Appendix: Figure ~\ref{fig:cost_evals_vs_params_faceted}.

\subsection{Experiment 4: Trainability comparisons across molecules as a function of input Hamiltonian prefix terms}
\begin{figure}[H]
\centering
\includegraphics[width=0.95\textwidth]{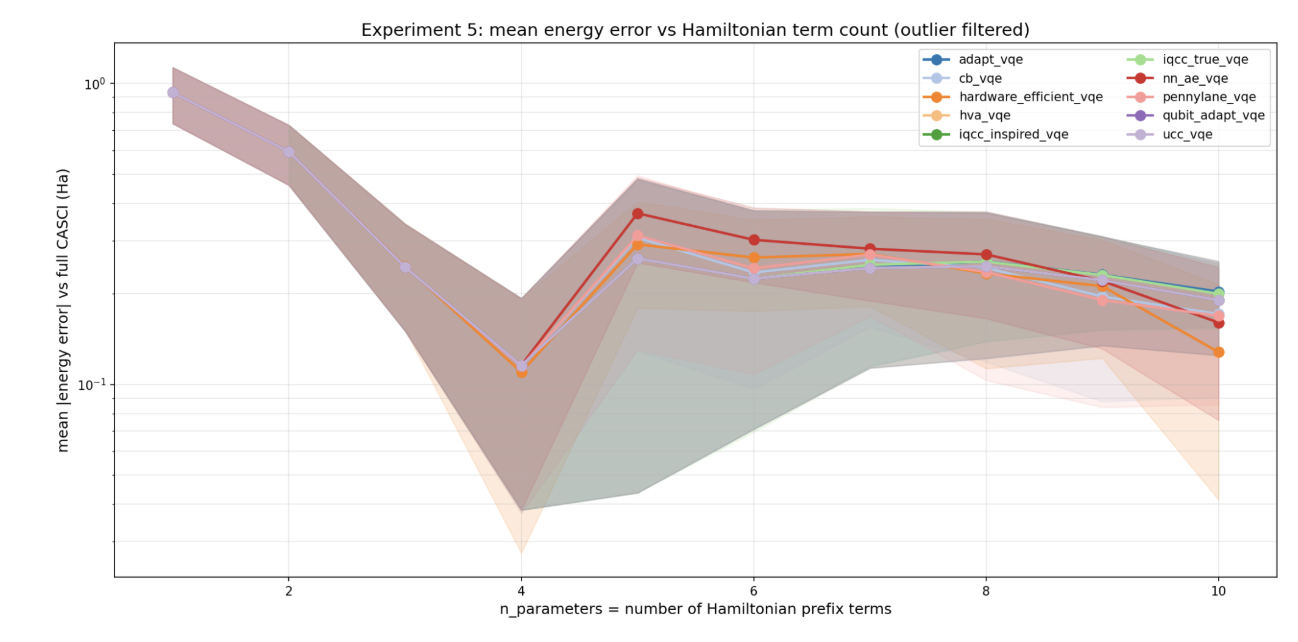}
\caption{Investigating how much keeping the first n\_parameters influences the mean energy deviation. For all algorithms, the error reaches a local minimum at $n_{parameters} = 4$, before decreasing from $n_{parameters} = 5$ to 10.}
\label{fig:mean-energy-error-ham}
\end{figure}

Another interesting question is to determine how much information is captured in the first few Hamiltonian prefix terms. We used the Hamiltonians directly from the QMProt dataset, but this experiment will also run on any other Hamiltonians that the users input. As can be seen in Figure ~\ref{fig:mean-energy-error-ham}, the mean energy error goes down from 1 to 4 first parameters (where the majority of the orbital information is presumably stored), and then jumps up again at 5, after which it undergoes a slow decrease until the first 10 parameters (after which experimentation was not run due to compute limitations). This may be because of the additional noise due to the additional parameters drowning out the marginal information from the additional parameters after $n_{parameters} = 4$.

\section{Discussion}
\subsection{Our Work}
We have created a repository for researchers to be able to run multiple VQE ansatz with multiple truncation methods in a fully automated pipeline. Researchers just need to install the requirements, and then they are able to run over 10 different ansatz, with automated active space truncation on their h5 files. This can be useful not only for biological molecules but also for other sorts of small to medium-sized molecules to better model near-term quantum hardware capabilities. In addition, researchers can run their own VQEs and benchmark them against other leading VQEs.
\newline
\newline
We have also investigated noise resilience for different molecules, as seen in Figures~\ref{fig:param-l2-drift} and~\ref{fig:l2-heatmap}, barren plateaus and initialization as in Figures~\ref{fig:part_b_convergence_summary_readable} and~\ref{fig:part_b_final_energy_delta_readable}, trainability (matched input Hamiltonian parameters) as in Figure~\ref{fig:trainability-fnevals}, and accuracy as a function of the number of parameters as in Figure~\ref{fig:mean-energy-error-ham}. These findings and benchmarking repository will be informational for researchers deciding tradeoffs on running molecular simulations on real quantum hardware.

\subsection{Limitations}
Previous research has shown that optimization algorithms can greatly influence the number of operations required for different VQE algorithms and methods. As such, it is possible that our results for the number of operations are overly dependent on the optimization algorithms that we are using for the VQE,~\cite{alfonso2025chemical,cheng2025variational,nicoli2025mlvqe}, and thus certain VQE ansatzes may perform better with different optimization methods.
\newline
\newline
Due to compute limitations, we only ran experiment 4 (energy error as a function of Hamiltonian prefix terms) until 10 prefix terms. However, noting the downward slope after n=5, it would be useful to extend it further.

\subsection{Potential Applications}
Benchmarking projects have been instrumental for quantum hardware development. To elaborate on hardware development, VQE has been used as a benchmark for NISQ devices. One primary example is the paper ''Quantum Chemistry as a Benchmark for Near-Term Quantum Computers'' by McCaskey et al. They used VQE, active space reduction, reduced unitary coupled cluster ansatz, and reduced density purification for error mitigation to ''provide a relevant baseline for future improvement of the underlying hardware, and a means for comparison across near-term hardware types''~\cite{mccaskey2019quantum}. This benchmarking repository may also serve as a benchmark for NISQ progress, especially as users can adjust noise levels and other parameters to determine what is required for different performance levels.
\newline
\newline
Researchers can also use the benchmark to conveniently test their own algorithms or even hardware limitations like noise against popular chemistry problems, like FeMoco~\cite{Li2019FeMocoElectronicComplexity} and caffeine.

\subsection{Future Research Directions}
Another interesting question is how much the choice of Jordan-Wigner for generating the Hamiltonians matters. For larger molecular systems, it would be useful to use the Bravyi-Kitaev method, which scales as O(logN) instead of O(N) as Jordan-Wigner does ~\cite{tilly2022vqe}. A potential research direction could be to see how transformation choice influences the scaling costs, and the degree to which it affects different VQE algorithms.

\section{Conclusion}
Our repository provides an extremely modular way for researchers to compare VQEs, and run different experiments for benchmarking, all specialized for larger biological molecules. Since amino acids are the molecular building blocks of proteins, the QMProt Hamiltonians provide a biologically meaningful intermediate scale between toy molecular systems and full protein active sites ~\cite{coronas2025qmprot}. Benchmarking VQE variants on these fragments can therefore inform which mixtures of ansatz, fermion-to-qubit mapping, active space selection, optimizer, and error-mitigation strategy are most suitable before extending to larger fragment based simulations, protein-ligand binding models, etc.

\section{Code Availability}

All reference code is available in the project repository.\footnote{\url{https://github.com/sanskriti-ss/qmprot_gse}}
\newline
\newline
The repository includes the instructions for downloading the QMProt Dataset, and how to use the repository to run different VQE on your own hamiltonian datasets.

Researchers are encouraged to make their own forks and open issues to contribute to the repository.

\section*{Acknowledgments}

The authors would like to thank many people who contributed to the completion of this project. Thank you to Professor Nicholas Mayhall from Indiana University who gave us invaluable feedback on our experiments. Thank you to Cody Fan and Nicolas Dirnegger, from the Quantum Computing Students Association (QCSA), for providing detailed help on everything from expanding our literature review to be more comprehensive, to giving feedback on paper structuring. In addition, thank you to Parfait Atchade-Adelomou, one of the QMProt Database authors, for his feedback on our progress.

\newpage
\section{Appendix}
\small

\begin{longtable}{p{2.0cm} p{4.2cm} p{8.0cm}}
\caption{Summary of VQE methods included in the benchmarking framework.}
\label{tab:vqe-methods}\\

\hline
VQE & Paper and Code & Description \\
\hline
\endfirsthead

\hline
VQE & Paper and Code & Description \\
\hline
\endhead

\hline
\endfoot

CB-VQE (PennyLane) 
& 
Paper: None written

Code: \url{https://www.pennylane.ai/qml/demos/tutorial_classically_boosted_vqe/}
& 
Enhances standard VQEs by performing a classical post-processing step that projects the molecular Hamiltonian onto a two-dimensional subspace spanned by the Hartree-Fock reference state and the VQE-optimized state. It then solves the generalized eigenvalue problem to guarantee a lower or equal energy compared to VQE, while requiring only Hadamard test circuits to compute cross-terms. This is valuable for near-term quantum devices where circuit depth and noise are limited. \\

\hline

NN-VQA
&
Paper: \url{https://doi.org/10.1103/PhysRevApplied.21.014053}

Code: \url{https://github.com/Miao-JQ/NN-VQA}
&
Uses neural-network-encoded variational quantum algorithms to address the challenges of implementing VQAs on NISQ devices. \\

\hline

CS-VQE
&
Paper: \url{https://quantum-journal.org/papers/q-2021-05-14-456/}

Code: \url{https://github.com/wmkirby1/ContextualSubspaceVQE}
&
A hybrid algorithm that reduces the number of qubits needed to achieve chemical accuracy. It works by mathematically dividing a molecule's energy calculation into a classical portion that is easily solved on a regular computer and a strictly quantum portion, called the contextual subspace. VQE is then run only on the contextual subspace. \\

\hline

iQCC
&
Paper: \url{https://pubs.acs.org/doi/10.1021/acs.jctc.9b01084}

Code: \url{https://github.com/sandbox-quantum/Tangelo-Examples/blob/main/examples/variational_methods/iqcc_using_clifford.ipynb}
&
Uses constant-depth, shallow quantum circuits that are iteratively updated through canonical transformations of the Hamiltonian on a classical computer. iQCC significantly reduces circuit depth, making it less prone to errors on NISQ devices. \\

\hline

VAns
&
Paper: \url{https://arxiv.org/abs/2103.06712}

Code: \url{https://github.com/matibilkis/qvans}
&
A variational ansatz architecture search method that adaptively builds quantum circuits by adding, removing, and optimizing gates. It aims to find compact, problem-suitable ansatz circuits rather than relying on a fixed ansatz structure. \\

\hline

Hamiltonian Variational Ansatz
&
Paper: None written

Code: \url{https://quantum.cloud.ibm.com/docs/en/api/qiskit/qiskit.circuit.library.hamiltonian_variational_ansatz}
&
A physics-informed ansatz that turns the structure of a molecular Hamiltonian into a quantum circuit by partitioning Hamiltonian terms into commuting groups and implementing time-evolution operators for each group. It mirrors the problem Hamiltonian, leading to fewer variational parameters while maintaining strong expressibility for ground-state preparation. This can reduce landscape complexity, improve trainability, and guide ansatz design. \\
\hline
PennyLane VQE
&
Paper: \cite{peruzzo2014variational}

Code: \url{https://pennylane.ai/qml/demos/tutorial_vqe/}
&
Implements a baseline VQE workflow in the PennyLane ecosystem. A parametrized circuit prepares a trial molecular state, the Hamiltonian expectation value is evaluated as the cost function, and a classical optimizer updates the circuit parameters to approximate the molecular ground-state energy. In our framework, this serves as a simple reference implementation for comparing more specialized ansatz constructions. \\

\hline

ADAPT-VQE
&
Paper: \cite{grimsley2019adaptive}

Code: Implementation adapted in our repository
&
Builds the ansatz adaptively rather than fixing the circuit structure in advance. At each outer-loop step, the algorithm evaluates gradients over an operator pool, selects the operator expected to recover the most correlation energy, appends it to the ansatz, and reoptimizes all variational parameters. This produces compact, molecule-tailored circuits but introduces additional gradient-screening and reoptimization overhead. \\

\hline

Qubit-ADAPT-VQE
&
Paper: \cite{tang2021qubitadapt}

Code: Implementation adapted in our repository
&
Uses the ADAPT-VQE growth strategy with a qubit-operator pool, typically based on Pauli-string generators rather than fermionic excitation operators. This makes the ansatz more hardware-oriented and can reduce circuit depth relative to fermionic ADAPT constructions, especially after qubit-space reductions where the original orbital structure is no longer directly meaningful. \\

\hline

Hardware-Efficient VQE
&
Paper: \cite{kandala2017hardware}

Code: Implementation adapted in our repository
&
Uses shallow, hardware-oriented layers of parametrized single-qubit rotations and entangling gates. In our implementation, the ansatz consists of repeated $R_y$ rotation layers and CNOT entangling layers. This design is easy to scale across qubit counts and is well suited for noise and barren-plateau diagnostics, but it is less chemically structured than UCC- or ADAPT-style ansätze. \\

\hline

iQCC-Inspired VQE
&
Paper: Inspired by iQCC, \cite{ryabinkin2020iqcc}

Code: Implementation adapted in our repository
&
Uses an ADAPT-style architecture with an operator pool motivated by iterative qubit coupled cluster methods. Unlike a full iQCC implementation, this variant should be described as repository-specific unless the final code can be traced to a published implementation. It is included to test whether iQCC-like qubit generators improve trainability or accuracy when used inside an adaptive VQE growth loop. \\

\hline
\end{longtable}
\normalsize

The number of active orbitals selected using a frontier orbital window in STO-3G (RHF, frozen core), consistent with common truncated CAS approaches in quantum simulated benchmarks. For our experiments, we used consistent rules (e.g. fitting for 6 orbitals.)
\newline
\newline
% A consistent rule here would be to select HOMO-1, HOMO, LUMO, LUMO+1, and additional π orbitals if present. 
% https://pubs.acs.org/doi/10.1021/acs.jctc.3c00123
(Note: The orbital counts below are properties of our default truncation strategy. For comparing different VQEs, we selected the 6 most relevant orbitals).
\begin{table}[h]
\centering
\begin{tabular}{lc}
\hline
Amino Acid & Active Orbitals \\
\hline
Alanine (Ala) & 4 \\
Arginine (Arg) & 6 \\
Asparagine (Asn) & 6 \\
Aspartate (Asp) & 6 \\
Cysteine (Cys) & 6 \\
Glutamate (Glu) & 6 \\
Glutamine (Gln) & 6 \\
Glycine (Gly) & 4 \\
Histidine (His) & 6 \\
Isoleucine (Ile) & 6 \\
Leucine (Leu) & 6 \\
Lysine (Lys) & 6 \\
Methionine (Met) & 6 \\
Phenylalanine (Phe) & 6 \\
Proline (Pro) & 6 \\
Serine (Ser) & 6 \\
Threonine (Thr) & 6 \\
Tryptophan (Trp) & 6 \\
Tyrosine (Tyr) & 6 \\
Valine (Val) & 6 \\
\hline
\end{tabular}
\caption{Active orbitals used for each amino acid.}
\label{tab:active_orbitals}
\end{table}

\begin{figure}[H]
\centering
\includegraphics[width=0.85\textwidth]{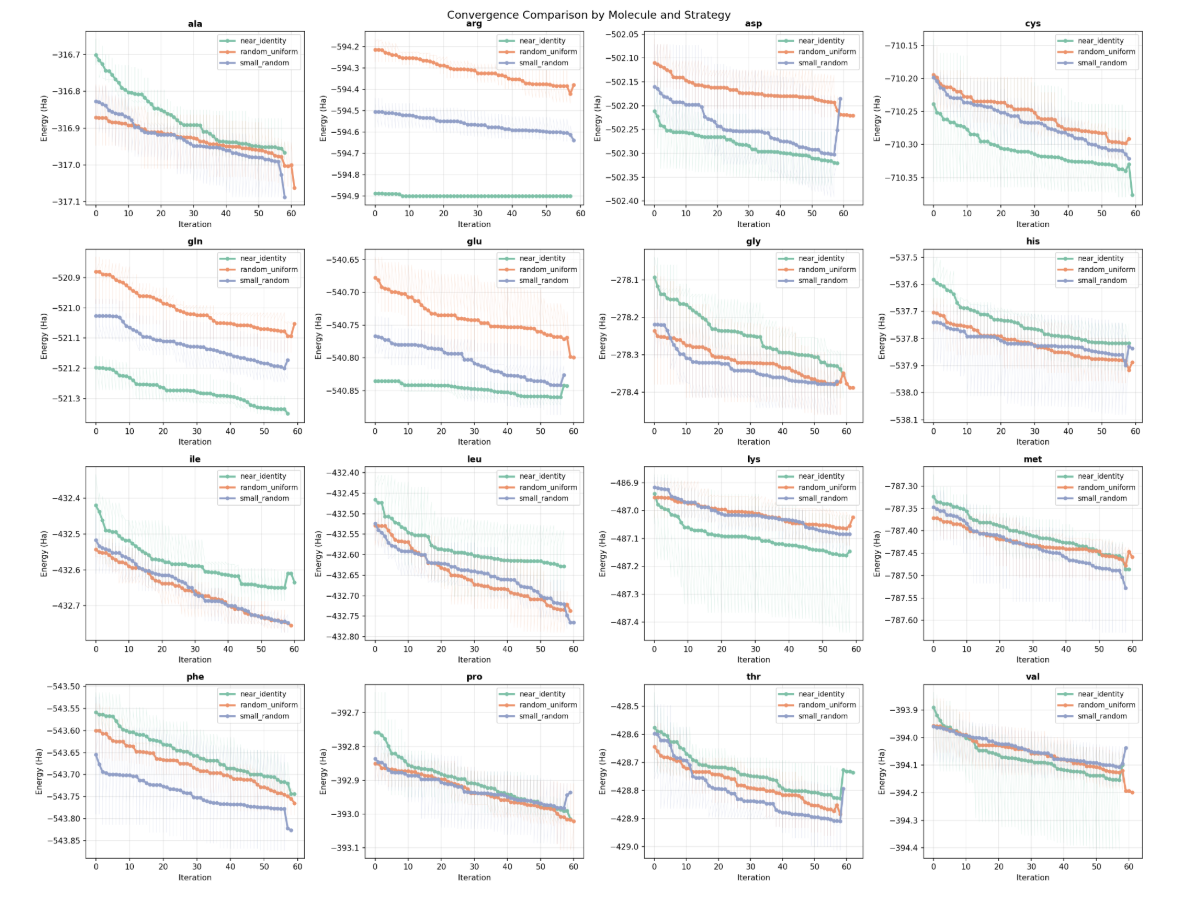}
\caption{Investigating convergence for different strategies for amino acids from the dataset, using \texttt{hardware\_efficient\_vqe}. This was run with COBYLA.}
\label{fig:molecule-convergence}
\end{figure}

\begin{figure}[H]
\centering
\includegraphics[width=0.85\textwidth]{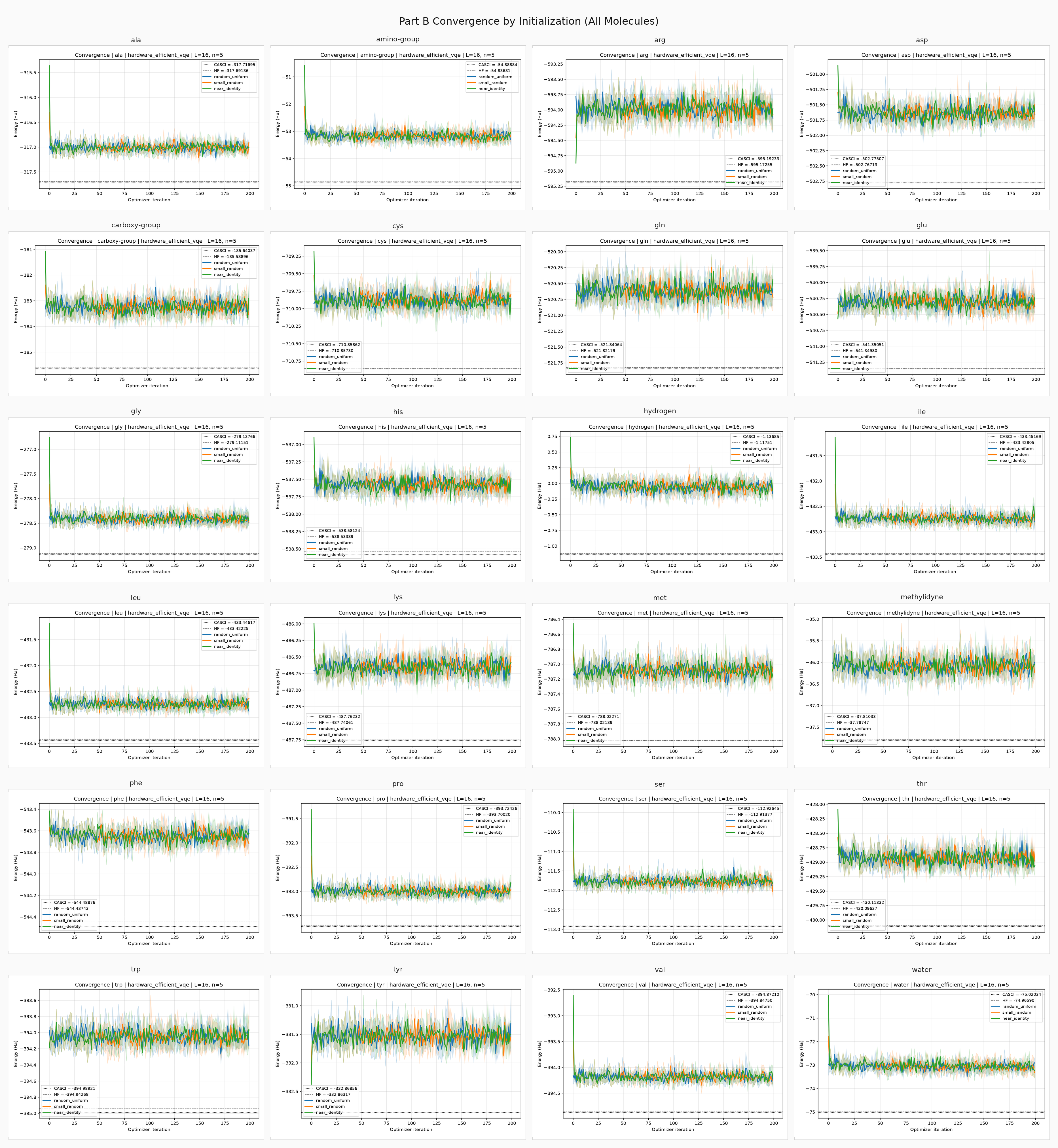}
\caption{Investigating convergence for different strategies for amino acids from the dataset, using \texttt{hardware\_efficient\_vqe}. Note that this was run with Bayesian Optimization, do not have any noise strengths in them, and more detailed compared to \ref{fig:part_b_convergence_summary_readable}}
\label{fig:convergence_by_init_all_molecules_grid}
\end{figure}

\begin{figure}[H]
\centering
\includegraphics[width=0.85\textwidth]{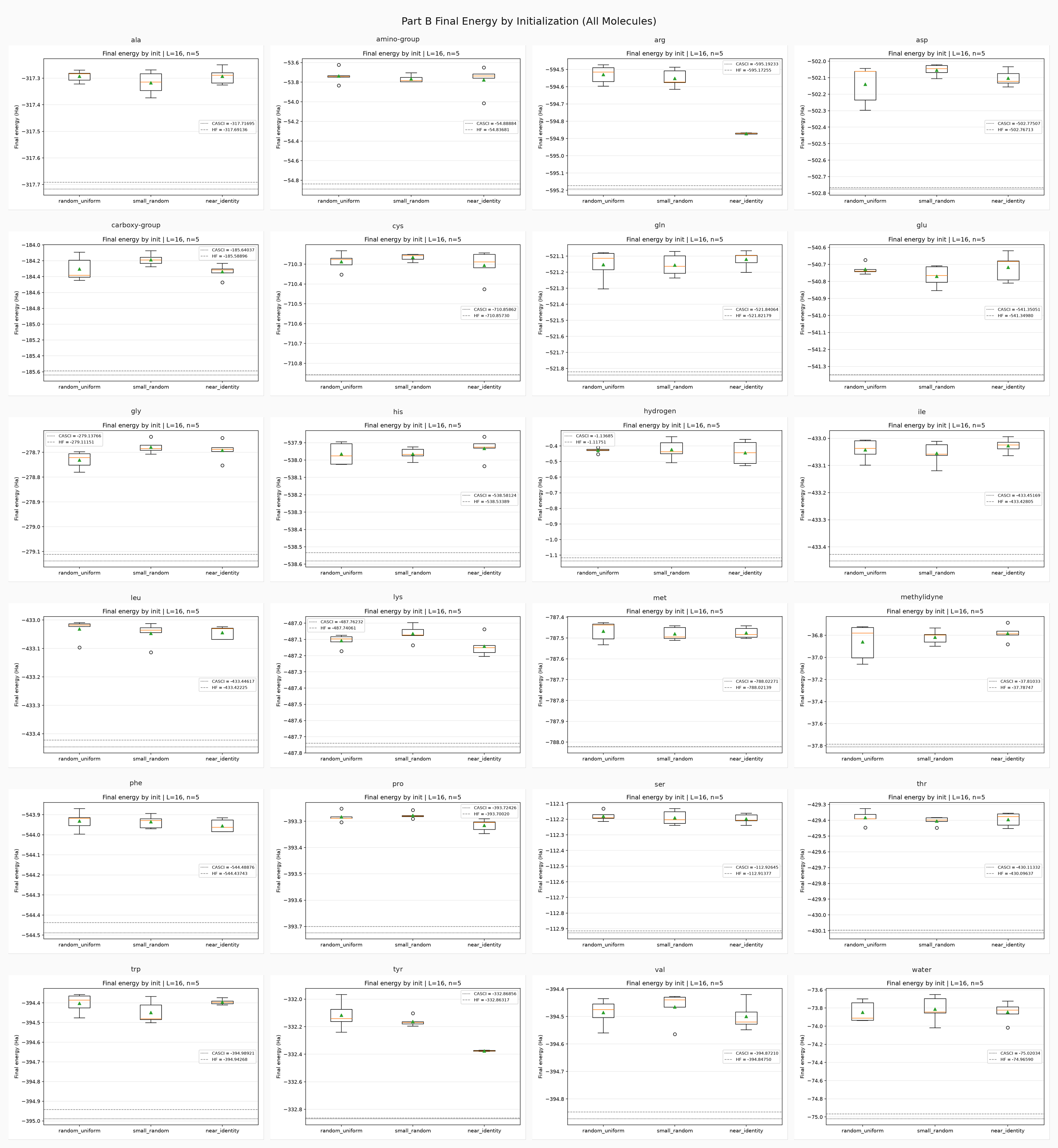}
\caption{This panel figure summarizes per-molecule final-energy-by-initialization comparisons in another way but more detailed compared to \ref{fig:part_b_final_energy_delta_readable} Note that these runs do not have any noise strengths in them.}
\label{fig:final_energy_by_init_all_molecules_grid}
\end{figure}

\begin{figure}[H]
\centering
\includegraphics[width=0.85\textwidth]{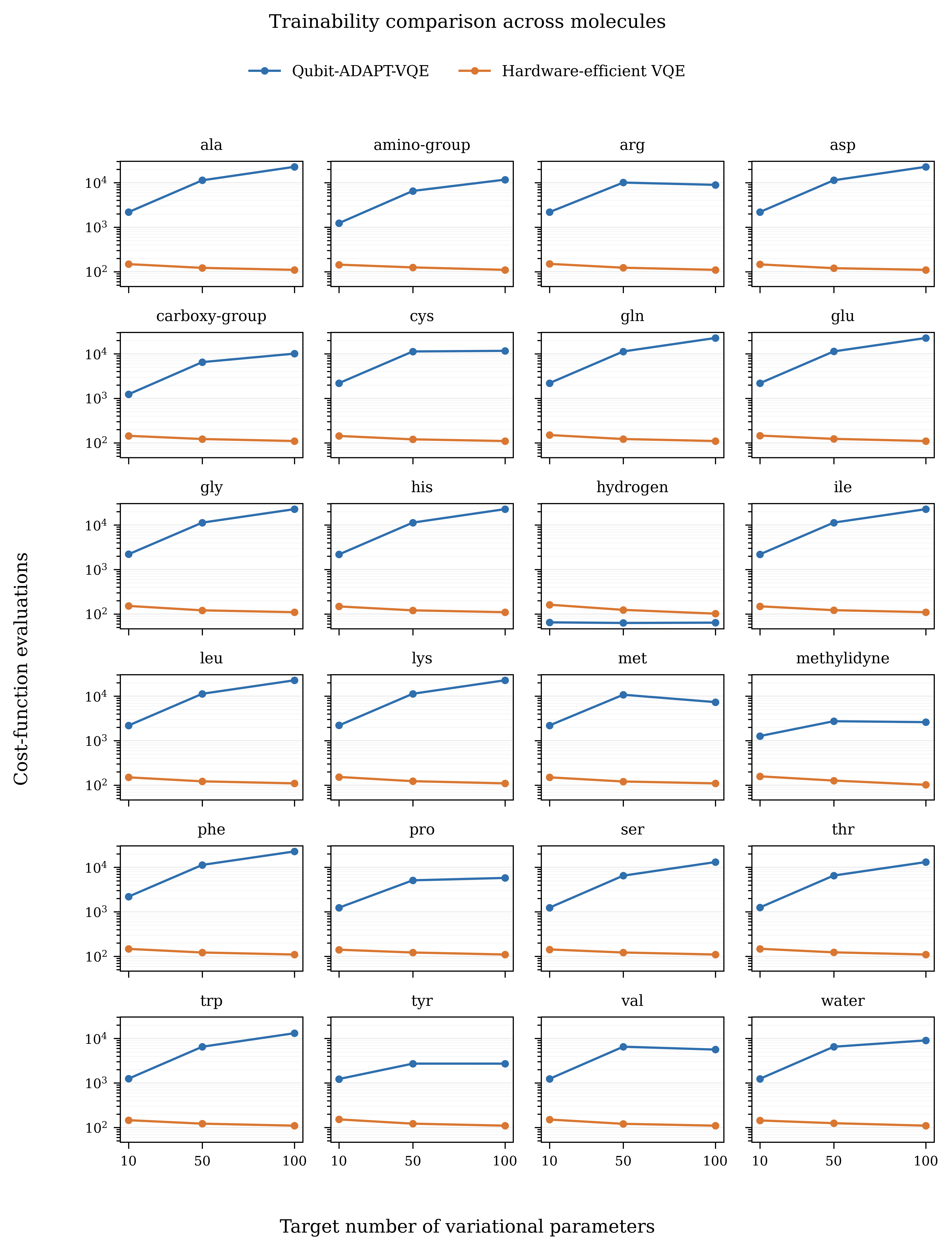}
\caption{Cost-function evaluations required by Qubit-ADAPT-VQE and hardware-efficient VQE across the trainability benchmark molecules. Each panel shows one molecule, with evaluations plotted on a logarithmic scale against the target number of variational parameters.}
\label{fig:cost_evals_vs_params_faceted}
\end{figure}

\begin{figure}[H]
\centering
\includegraphics[width=0.85\textwidth]{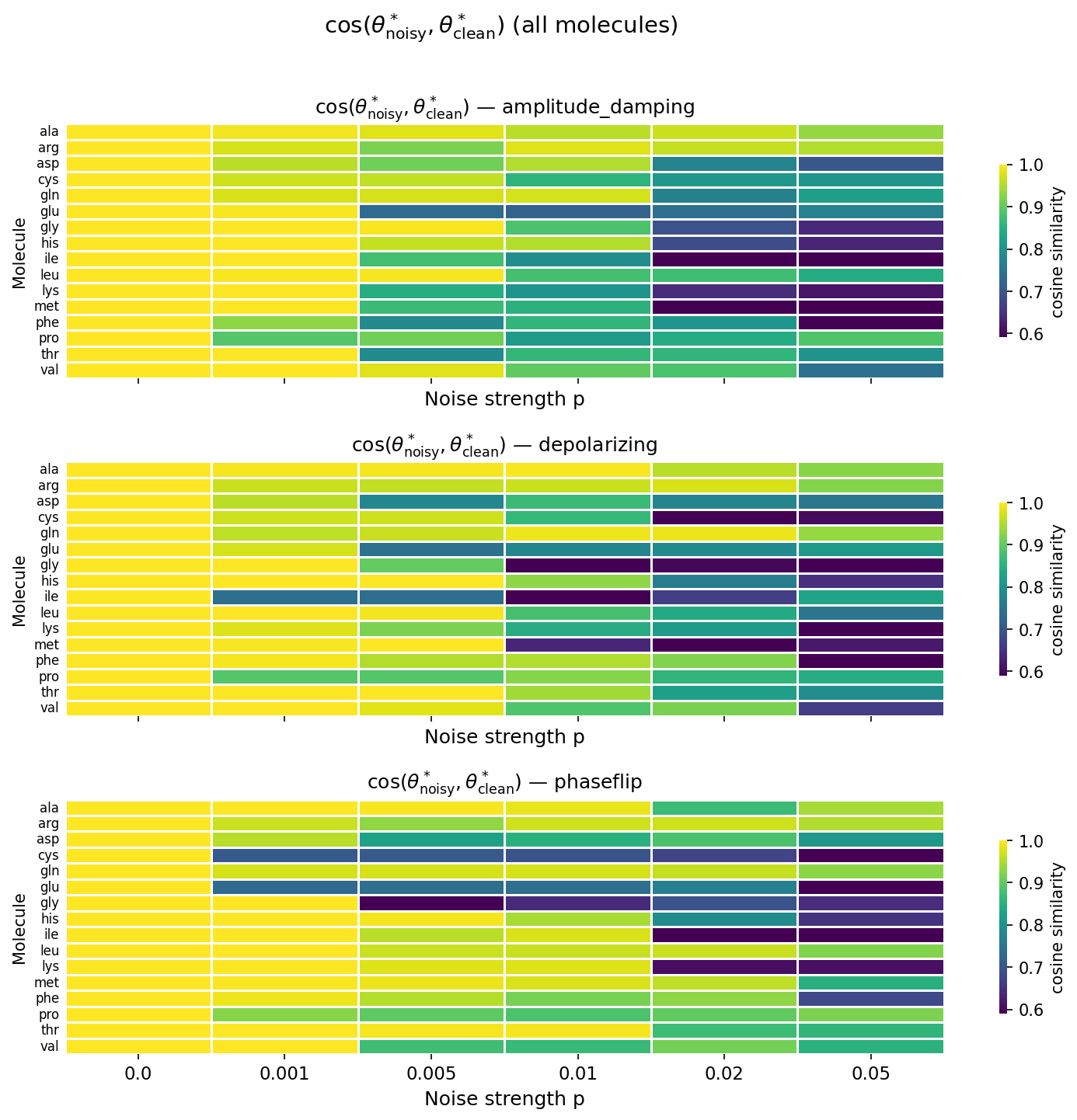}
\caption{Cosine similarity heatmap, with the same conditions as Figure ~\ref{fig:l2-heatmap}}
\label{fig:heatmap_param_drift_cosine}
\end{figure}

\clearpage
\bibliographystyle{unsrtnat}
\bibliography{references}

\end{document}